\documentclass[journal]{IEEEtran}

\usepackage{amsfonts}
\usepackage{tikz}
\usepackage{graphicx} 
\usepackage{epsfig} 
\usepackage{amsthm}
\usepackage{mathptmx} 
\usepackage{color}
\usepackage{amsmath} 
\usepackage{mathdots}
\usepackage{amssymb}  
\usepackage{multicol}
\usepackage{multirow}
\usepackage{eucal}
\usepackage{mathtools, cuted}
\usepackage{mathtools}
\usepackage{gensymb}
\usepackage{nopageno}
\usepackage{algorithm}
\usepackage[noend]{algpseudocode}
\usepackage{algorithmicx,algorithm}
\usepackage{mathrsfs}
\usepackage{bm}
\definecolor{LightGray}{rgb}{0.75,0.75,0.75}

\newtheorem{theorem}{Theorem}
\newtheorem{itlemma}{Lemma}
\newtheorem{itdefinition}{Definition}
\newtheorem{itproposition}{Proposition}
\newtheorem{itremark}{Remark}

\newtheorem{itexample}{Example}
\newtheorem{itproblem}{Problem}
\newtheorem{itassumption}{Assumption}

\newenvironment{definition}{\begin{itdefinition}\rm}{\end{itdefinition}}
\newenvironment{remark}{\begin{itremark}\rm}{\end{itremark}}
\newenvironment{lemma}{\begin{itlemma}\rm}{\end{itlemma}}

\newenvironment{assumption}{\begin{itassumption}\rm}{\end{itassumption}}
\newenvironment{proposition}{\begin{itproposition}\rm}{\end{itproposition}}
\DeclareMathAlphabet{\mathpzc}{OT1}{pzc}{m}{it}

\usepackage{xparse}
\NewDocumentCommand{\INTERVALINNARDS}{ m m }{
    #1 {,} #2
}
\NewDocumentCommand{\interval}{ s m >{\SplitArgument{1}{,}}m m o }{
    \IfBooleanTF{#1}{
        \left#2 \INTERVALINNARDS #3 \right#4
    }{
        \IfValueTF{#5}{
            #5{#2} \INTERVALINNARDS #3 #5{#4}
        }{
            #2 \INTERVALINNARDS #3 #4
        }
    }
}
\makeatletter
\def\endthebibliography{%
  \def\@noitemerr{\@latex@warning{Empty `thebibliography' environment}}%
  \endlist
}
\makeatother

\begin{document}
\title{A Separation-Based Methodology to Consensus Tracking of Switched High-Order Nonlinear Multi-Agent Systems}

\author{\IEEEauthorblockN{Maolong Lv, Wenwu Yu, Jinde Cao, Simone Baldi}
\thanks{This work is partially supported by Descartes Fellowship (French + Netherlands project) (\emph{Corresponding author: Simone Baldi}).

M. Lv is with the Delft Center for Systems and Control, Delft
University of Technology, Mekelweg 2, Delft 2628 CD, The Netherlands (email:
M.Lyu@tudelft.nl).

W. Yu and J. Cao are with the Department of Mathematics, Southeast University, Nanjing 210096, China (e-mail: wwyu@seu.edu.cn; jdcao@seu.edu.cn).

S. Baldi is with the Department of Mathematics, Southeast University,
Nanjing 210096, China, and with the Delft Center for Systems and Control, Delft
University of Technology, Mekelweg 2, Delft 2628 CD, The Netherlands (email: S.Baldi@tudelft.nl).
}}%

\maketitle

\begin{abstract}
This work investigates a reduced-complexity adaptive methodology to consensus tracking for a team of uncertain high-order nonlinear systems with switched (possibly asynchronous) dynamics. It is well known that high-order nonlinear systems are intrinsically challenging as feedback linearization and backstepping methods successfully developed for low-order systems fail to work. At the same time, even the adding-one-power-integrator methodology, well explored for the single-agent high-order case, presents some complexity issues and is unsuited for distributed control. At the core of the proposed distributed methodology is a newly proposed definition for separable functions: this definition allows the formulation of  a separation-based lemma to handle the high-order terms with reduced complexity in the control design. Complexity is reduced in a twofold sense: the control gain of each virtual control law does not have to be incorporated in the next virtual control law iteratively, thus leading to a simpler expression of the control laws; the order of the virtual control gains increases only proportionally (rather than exponentially) with the order of the systems, dramatically reducing high-gain issues.
\end{abstract}


\IEEEpeerreviewmaketitle

\section{Introduction}
Distributed leaderless or leader-following consensus control of nonlinear multi-agent systems is more challenging but also potentially more applicable than its linear counterpart. Similarly to the linear case, the goal is to steer a team of agents to a not globally known trajectory using only locally available information collected from neighboring agents[1-3]. In recent years, leaderless or leader-following consensus results have been obtained for two large families of nonlinear multi-agent systems: strict-feedback [4-21] and pure-feedback multi-agent systems [22-25]. For these families, the commonly adopted approach is an extension of the well-known backstepping technique [26] in a distributed sense. When the nonlinear functions are unknown, approximators such as neural networks and fuzzy logic systems have been incorporated in such a design. Switching dynamics can also be handled via the common Lyapunov function method [7], [9], [23], [25]. Although strict-feedback and pure-feedback systems are among the most studied dynamics in the nonlinear control field, there exist extensions to these dynamics: most notably, high-order nonlinear systems are a generalization of strict-feedback or pure-feedback systems in which integrators with positive odd powers will appear in the dynamics (chain of positive odd power integrators). On the contrary, strict-feedback or pure-feedback systems contain chains of integrator whose powers are equal to one (linear integrators). Literature has shown that high-order dynamics, appearing in aerospace and robotic applications [27]-[33], are extremely challenging to deal with, as their linearized dynamics might possess uncontrollable modes whose eigenvalues are on the right half-plane [34], making all standard methodologies in [4]-[25] fail. Let us remark that the term "high-order" used in [27]-[33] is different than the term "high-order" used in [12], [13], [15]-[16]: the former term is often used in the nonlinear control community to indicate that the integrators in the chain  may not only have power equal to one (low-order dynamics), but higher or equal to one (high-order dynamics); the latter term is often used in the consensus community to indicate that the chain is not composed by one integrator with power equal to one (first-order agents), nor by two integrators with power equal to one (second-order agents), but by more than two integrators with power equal to one (high-order agents). Therefore, from the point of view of [27]-[33], the dynamics of [12], [13], [15]-[16] are strict-feedback and in fact standard backstepping methods have been successfully adopted there. In this work, the term "high-order" is to be intended in the sense of [27]-[33] (chain of positive odd power integrators), for which no standard backstepping technique can be adopted.

In place of the standard backstepping, the adding-one-power-integrator technique was successfully proposed in [28] to handle high-order dynamics. Progress made for the single high-order system case include relaxing the growth condition on the nonlinear functions [29]-[30], [34] and employing neural network or fuzzy logic approximators to handle completely unknown nonlinearities [31]-[33], [35]. However, it has to be emphasized that a direct extension of the standard adding-one-power-integrator technique in a distributed sense is not meaningful due to some complex aspects of the procedure. At least the following two complex aspects are worth mentioning: (a) the high-power terms are separated from the control gain functions via \emph{separation lemmas} that make the order of the virtual control gains grow exponentially with the order of the system; (b) the control gain of each virtual control is incorporated into the next virtual control law iteratively, thus increasing the control complexity at each step. Such issues result in high-complexity and high-gain designs which might be prohibitive for multi-agent systems with low computational power and limited actuation. Therefore, the crucial open question motivating this research is \emph{how can reduced-complexity distributed methodologies be designed for high-order nonlinear multi-agent systems?}

The main contribution of this work is to answer this question for a large class of uncertain high-order nonlinear multi-agent systems, which can exhibit heterogeneous nonlinearities and switched  dynamics with possibly asynchronous switches among the agents. At the core of the proposed methodology is a newly proposed definition for separable functions and a new separation-based lemma to deal with the high-power terms. The lemma decreases the complexity of the distributed consensus design in a twofold direction: it avoids incorporating the control gain of each virtual control in the next virtual control law, thus sensibly reducing the complexity of the control action; it allows the order of the control gains to increase only proportionally (rather than exponentially) with the order of the systems, thus dramatically reducing any high-gain issue (cf. the discussions in Remark 4 and 5 of this manuscript).

\emph{Notations:} The notations adopted in this paper are standard: $\mathbb R$ and $\mathbb R^n$ denote the set of real numbers and the $n$-dimensional Euclidean space, respectively. $Q_{\mathrm{odd}}$ represents the set of positive odd integers. $\|\cdot\|$ refers to either the Euclidean vector norm or the induced matrix 2-norm. For compactness and whenever unambiguous, throughout this paper, some variable dependencies might be dropped, e.g. $\xi$, $h_{f,k}^j$, $\ell_{f,k}$ and $\upsilon_{f,k}$ can be used to denote $\xi(x_1,x_2)$, $h_{f,k}^j(\overline {x}_{f,k})$, $\ell_{f,k}(s_{f,k+1},v_{f,k})$, and $\upsilon_{f,k}(s_{f,k+1},v_{f,k})$, respectively.
\section{Problem Formulation and Preliminaries}
Let us first give some preliminaries on graph theory. The communication topology is described by a directed graph $\mathscr G\triangleq(\mathscr V, \mathscr E)$, with $\mathscr{V} \triangleq \{0,1,\ldots, N\}$ being the set of nodes (agents) and with $\mathscr E\subseteq\mathscr V\times \mathscr V$ being the set of directed edges between two distinct agents (self-edges are not allowed). A directed edge $(j,i)\in \mathscr E$ represents that agent $i$ can obtain information from agent $j$. The neighbor set of agent $i$ is denoted by $\mathscr N_i=\{j|(j,i)\in \mathscr E\}$: this is the set of agents from which agent $i$ can obtain information. We reserve index 0 to the so-called leader agent: because agent 0 plays a special role, let us consider the subgraph defined by $\overline {\mathscr G}\triangleq\big(\overline{\mathscr V},\overline{\mathscr E}\big)$ with $\overline{\mathscr {V}}\triangleq \{1,2\ldots, N\}$ and $\overline{\mathscr E}$  defined accordingly. For this subgraph, let us define the connectivity matrix $\overline{\mathscr A}=[a_{ij}]\in \mathbb R^{N\times N}$: if $(j,i)_{i\neq j}\in \overline {\mathscr E}$, then $a_{ij}=1$, otherwise $a_{ij}=0$ (note that $a_{ii}=0$). The Laplacian matrix $\mathscr L$ associated with $\mathscr G$ is defined as
\begin{align*}
\mathscr L=\left[ \begin{array}{cc}0 & \textbf{0}_{1\times N} \\
 -\mu & \overline {\mathscr L}+\mathscr B\\
\end{array}
\right]
\end{align*}
 with $\mu=[\mu_1,\ldots,\mu_N]^{\mathrm T}$, being $\mu_i = 1$ if the leader $0\in \mathscr N_i$, and $\mu_i=0$ otherwise. Also, $\mathscr B=\mathrm{diag}[\mu_1,\ldots,\mu_N]^T$ and $\overline {\mathscr L}=\overline {\mathscr D}-\overline {\mathscr A}$ is the Laplacian matrix related to $\overline {\mathscr G}$ with $\overline {\mathscr D}=\mathrm{diag}[d_1,\ldots,d_N]$, where $d_i=\sum_{j\in \mathscr N_i}a_{ij}$.

Consider a team of $N$ $(N\ge 2)$ switched high-order nonlinear multi-agent systems whose dynamics are given by
\begin{equation}\label{systemmodel}
\left\{\begin{split}
  \dot {x}_{f,k}&=\varphi_{f,k}^{\sigma_f(t)}(\overline {x}_{f,k})+h_{f,k}^{\sigma_f(t)}(\overline {x}_{f,k})x_{f,k+1}^{r_{f,k}},\\
        \dot {x}_{f,n_f}&=\varphi_{f,n_f}^{\sigma_f(t)}(\overline {x}_{f,n_f})+h_{f,n_f}^{\sigma_f(t)}(\overline {x}_{f,n_f})u_{f}^{r_{f,n_f}},\\
  y_f&=x_{f,1},
\end{split}\right.
\end{equation}
with $1\le f\le N$, $1\le k\le n_f-1$, $\overline {x}_{f,k}=[x_{f,1},\ldots,x_{f,k}]^T\in \mathbb R^k$. The subscript $f$ stands for ``follower", in order to distinguish them from the leader agent, as clarified later. In (1), $\sigma_f(\cdot)$: $[0,+\infty)\rightarrow \mathpzc{M}_f=\{1,2,...,m_f\}$ is the switching signal for the $f$th follower, with $\mathpzc M_f$ denoting the switching mode set and $m_f$ denoting the number of modes for the $f$th follower; $r_{f,k}\in Q_{\mathrm{odd}}$ are the high powers (positive odd integers), and $u_f^j \in\mathbb R$ is the control input for the $j$th mode of the $f$th follower. For each mode $\sigma_{f}(t)$, the functions $\varphi_{f,k}^{\sigma_f(t)}(\cdot)$ and $h_{f,k}^{\sigma_f(t)}(\cdot)$ are unknown continuous functions. The following remarks highlight the difference between (1) and other multi-agent system models considered in literature.
\begin{remark} (Novelty of the class) The multi-agent models in [4]-[21] are strict-feedback low-order, i.e. special cases of (1) when all the powers $r_{f,k}$ are equal to one. Apart from this, (1) also possesses several levels of heterogeneity because: each follower agent exhibits its own switching $\sigma_f(\cdot)$, leading to possible asynchronous switching among the $N$ followers; each follower agent has its own state dimension $n_f$; the unknown switched nonlinearities $\varphi_{f,k}^{\sigma_f(t)}(\cdot)$ and $h_{f,k}^{\sigma_f(t)}(\cdot)$ are heterogeneous, i.e. possibly different for each follower. While similar levels of heterogeneity are considered in the pure-feedback multi-agent models in [22]-[25], those multi-agent systems models are also homologous to the strict-feedback low-order case, i.e. they can be equivalently transformed into the strict-feedback low-order form using the mean-value theorem.
\end{remark}

\begin{remark}(relevance of high-order nonlinear dynamics)
It is crucial to underline the importance of the high-order nonlinear case from both a mathematical and an engineering perspective: from a mathematical point of view, standard feedback linearization and backstepping methods can be successfully used for low-order systems, but they do not work anymore for (1) due to the fact that the linearized dynamics may have uncontrollable modes [28], [34]. From an engineering point of view, dynamics (1) can describe a large class of underactuated, weakly coupled, mechanical systems [34], [36]. In particular, positive odd powers naturally appear in flexible structures to describe anti-symmetric restoring forces (spring forces) [34], [36].
\end{remark}

To facilitate distributed control design for (\ref{systemmodel}), the following standard assumptions are made.
\begin{assumption}[23] \label{assumpMatchCond}
The leader agent  0 is represented by a leader output signal $y_r$, which is continuous, bounded, and available only to a subset of the follower agents. Furthermore, $\dot y_r$ is bounded and not available to any follower agent. The bounds for $y_r$ and $\dot y_r$ are unknown.
\end{assumption}

\begin{assumption}[20] \label{assumpDelay}
The multi-agent communication is represented by a directed graph $\mathscr G=(\mathscr V, \mathscr E)$, which contains at least one directed spanning tree with the leader agent as the root.
\end{assumption}

\begin{assumption}[33] \label{assumpMatchCond}
For each follower agent $f$, we assume the sign of $h_{f,k}^j$ is positive and there exist known real positive constants $\overline h_{f,k}^{j}$ and $\underline h_{f,k}^{j}$, $(1\le k\le n_f,~j\in\mathpzc M_f)$ such that $\underline h_{f,k}^{j}\le h_{f,k}^{j}(\cdot)\le \overline h_{f,k}^j$.
\end{assumption}

\begin{remark}(meaning of assumptions)
Assumption 1 implies that the leader information is only available to a small fraction of followers. Assumption 2 implies that $\overline {\mathscr L}+\mathscr B$ is a nonsingular  $\mathpzc M$-$\mathrm {matrix}$$\footnote{An $\mathpzc M$-matrix is a square matrix with non-positive off-diagonal entries and non-negative principal minors.}$ and guarantees the feasibility of consensus [37]. Assumption 3 is a general controllability condition for many classes of nonlinear dynamics, including strict-feedback, pure-feedback and high-order nonlinear systems [31]-[33].
\end{remark}
\subsection{Technical lemmas}
The following lemmas are useful for deriving the main results.
\begin{lemma} \label{02}
[29] For any $x_1\in \mathbb R$ and $x_2\in\mathbb R$, and given positive integers $b_1$, $b_2$ and any real-valued function $\xi(x_1,x_2)>0$, it holds that
\begin{equation}\label{}
  |x_1|^{b_1}|x_2|^{b_2}\le \frac{b_1\xi|x_1|^{b_1+b_2}}{b_1+b_2}+\frac{b_2\xi^{-\frac{b_1}{b_2}}|x_2|^{b_1+b_2}}{b_1+b_2}.
\end{equation}
\end{lemma}

\begin{lemma}
[33] Let $x_1$ and $x_2$ be real-valued functions. There exist a positive odd integer $\bar h$ and a constant $ \bar\lambda\ge 1$ such that \begin{subequations}
\begin{align}\label{03}
  \Big|x_1^{ \bar {h}}-x_2^{ \bar {h}}\Big| &\le  \bar {h}\big|x_1-x_2\big|\Big|x_1^{\bar {h}-1}+x_2^{ \bar {h}-1}\Big| \\
    |x_1+x_2|^ {\bar {\lambda}} & \le 2^{\bar { \lambda}-1}\big(|x_1|^ {\bar {\lambda}} +|x_2|^{\bar {\lambda}} \big).
\end{align}
\end{subequations}
\end{lemma}
The following definition, lemma and proposition are introduced to the purpose of reduced-complexity control, as it will be remarked later (cf. Remarks 4 and 5).
\begin{definition}
For any $x_1\in \mathbb R$, $x_2\in\mathbb R$, the function $\digamma(\cdot)$: $\mathbb R\rightarrow \mathbb R$ is said to be a \emph{separable function} provided that the following is satisfied:
\begin{equation}
  \digamma(x_1+x_2)=\ell(x_1,x_2)\digamma(x_1)+\upsilon(x_1,x_2)\digamma(x_2),
\end{equation}
where $\ell(x_1,x_2)\in\big[\underline \ell_1,\overline \ell_1\big]$ with $\underline \ell_1=1-d$ and $\overline \ell_1=1+d$, with $d\in(0,1)$ a constant, $|\upsilon(x_1,x_2)|\le \overline\upsilon(d)$ with $\overline\upsilon(d)$ denoting a positive continuous function, independent of $x_1$ and $x_2$.
\end{definition}

\begin{proposition}
For any $x_1\in \mathbb R$, $x_2\in\mathbb R$, the function $\digamma(\cdot)$ is a separable function if the following hold:\\
(i)  $\digamma(x_1x_2)=\digamma(x_1)\digamma(x_2)$\\
(ii)  For $p\in \mathbb R$ and $0<d<1$, a positive continuous function $\overline\upsilon(d)$ exists satisfying $|\digamma(\overline p)-1|\le \overline\upsilon(d)|\digamma(p)|+d$, where $\overline p=p+1$.
\end{proposition}
PROOF. See Appendix.
\begin{lemma}
A function $\digamma(z)=z^r$ with $r$ being a positive odd integer is a separable function.
\end{lemma}
PROOF. See Appendix.
\subsection{Consensus problem}
Define the tracking error for the $f$-th follower as
  \begin{align}\label{05}
    s_{f,1} =\sum_{l\in \mathscr N_f}a_{fl}(y_f-y_{l})+\mu_f(y_f-y_r),
  \end{align}
where $f=1,\ldots,N$. After defining $s_1=[s_{1,1},\ldots,s_{N,1}]^T\in \mathbb R^N$, one has $s_1=(\overline{ \mathscr L}+\mathscr B)\delta$ where $\delta=\bar y-\bar {y}_r$ with $\bar y=[y_1,\ldots,y_N]^T$ and $\bar {y}_r=[y_r,\ldots,y_r]^T$. Due to the nonsingularity of $\overline{\mathscr L}+\mathscr B$, it holds that $\|\delta\|\le \frac{\|s_1\|}{\underline{\lambda}_{\mathrm{min}}\big(\overline{\mathscr L}+\mathscr B\big)}$ [37], being $\underline{\lambda}_{\mathrm{min}}$ the minimum singular value of $\overline{\mathscr L}+\mathscr B$.

The following definition delineates the problem formulation, i.e. the desirable properties for the tracking errors in (5).
\begin{definition} [23] The consensus tracking errors (5) are said to be cooperatively semi-globally asymptotically bounded if there exist tunable constants $c_1>0$, $c_2>0$ and bounds $\beta_1>0$, $\beta_2>0$, such that for every $\alpha_1\in(0,c_1)$ and $\alpha_2\in(0,c_2)$, then $\|y_f(t_0)-y_r(t_0)\|\le \alpha_1\Rightarrow\|y_f(t)-y_r(t)\|\le \beta_1$ and $\|y_f(t_0)-y_{l}(t_0)\|\le \alpha_2\Rightarrow\|y_f(t)-y_{l}(t)\|\le \beta_2$ as $t\rightarrow \infty$ where $t_0$ is an initial time, $f=1,\ldots,N$, $l=1,\ldots,N$, and $f\neq l$.
\end{definition}

It is worth noticing that a bounded consensus result is sought, for the reason that asymptotic results are in general impossible (even locally) for high-order dynamics [27]. Therefore, asymptotic consensus results are in general impossible for high-order dynamics. In the following section we will discuss the design achieving the result of Definition 2.
\section{Proposed distributed consensus design}
Let us define the following variables for the $f$-th follower
\begin{equation}\label{}
   s_{f,k}  =x_{f,k}-v_{f,k-1},~~k=2,\ldots,n_f,
\end{equation}
and let us propose the following design, whose rationale will be given in Sect. III-A.
 \begin{align}\label{}
   v_{f,1} &=-s_{f,1}\underbrace{\Im_{f,1}^{\frac{1}{r_{f,1}}}\bigg(c_{f,1}+\zeta_{f,1}^{{\overline {r}_{f,1}}}\widehat{\Xi}_{f,1}\Theta_{f,1}^{{\overline {r}_{f,1}}}+b_{f,1}^{{\overline {r}_{f,1}}}\bigg)^{{\frac{1}{r_{f,1}}}}}_{\varsigma_{f,1}}, \\
  \Im_{f,1} &=\Big[{\underline {h}_{f,1}}(d_f+\mu_f)(1-d)\Big]^{-1},\\
    v_{f,k} &=-s_{f,k}\underbrace{\Im_{f,k}^{\frac{1}{r_{f,k}}}\bigg(c_{f,k}+\zeta_{f,k}^{{\overline {r}_{f,k}}}\widehat{\Xi}_{f,k}\Theta_{f,k}^{{{\overline r_{f,k}}}}+b_{f,k}^{{{\overline {r}_{f,k}}}}\bigg)^{{\frac{1}{r_{f,k}}}}}_{\varsigma_{f,k}}, \\
   \Im_{f,k}&=\Big[{\underline {h}_{f,k}}(1-d)\Big]^{-1},~(k=2,\ldots,n_f),\\
   u_f&\triangleq u_f^j=v_{f,n_f},~~j\in\mathpzc M_f,
   \end{align}
where $r_f=\underset{1\le k\le n_f}{\mathrm{max}}\big\{r_{f,k}\big\}$, $\overline {r}_{f,k}=\frac{r_f+1}{r_f-r_{f,k}+1}$, $\underline {r}_{f,k}=\frac{r_f+1}{r_{f,k}}$, ${\underline {h}_{f,k}=\min\big\{\underline {h}_{f,k}^{j},j\in\mathpzc M_f\big\}}$, $\zeta_{f,k}>0$, $b_{f,k}>0$ and $c_{f,k}>0,(k=1,\ldots,n_f)$ are design constants.

Further, the parameters $\widehat{\Xi}_{f,k}$, $k = 1, ..., n_f$, are adapted via the laws
\begin{equation}\label{adaptationlaw}
  \dot {\widehat \Xi}_{f,k}=\beta_{f,k}\zeta_{f,k}^{\overline {r}_{f,k}}s_{f,k}^{r_f+1}\Theta_{f,k}^{\overline r_{f,k}}-\beta_{f,k}\sigma_{f,k}\widehat {\Xi}_{f,k},
\end{equation}
where $\beta_{f,k}>0$ denotes a tuning rate and $\sigma_{f,k}>0$ is a design parameter. The adaptation law (12) arises from the use of linear-in-the-parameter approximators for the nonlinear dynamics in (1), e.g. as done in high-order literature [31]-[33]. However, differently from this literature, the design of (7)-(11) avoids high-complexity and high-gain aspects (cf. discussion in Remark 5).
 The design of (7)-(12), as well as the remaining parameters $\Theta_{f,k}$, $(k=1,\ldots,n_f)$ shall be specified in the following design steps.

\subsection{Design Steps}
\textbf{Step $f,1~(f=1,\ldots,N)$ : } The time derivative of $s_{f,1}$ along (\ref{systemmodel}) and (5) is
\begin{equation}\label{}
  \dot {s}_{f,1}=(d_f+\mu_f)h_{f,1}^j(x_{f,1})x_{f,2}^{r_{f,1}}+H_{f,1}^j,
\end{equation}
where $H^j_{f,1}$ is a function defined as
 \begin{align}\label{}
H_{f,1}^j=&(d_f+\mu_f)\varphi_{f,1}^j(x_{f,1})
-\sum_{l\in \mathscr {N}_f}a_{fl}\bigg(\varphi_{l,1}^j(x_{l,1})\notag\\
&+h_{l,1}^j(x_{l,1})x_{l,2}^{r_{l,1}}\bigg)-\mu_f\dot {y}_r(t).
 \end{align}
 From Assumptions 1 and 3, and along similar ideas to [23], [33], [38], one can conclude that there exist a continuous function $F_{f,1}(Z_{f,1})$ and a linear-in-the-parameter approximator $\widehat F_{f,1}\big(Z_{f,1}\big|{W}_{f,1}^{*}\big)$ such that, for any $j\in \mathpzc M_f$,
 \begin{align*}
&s_{f,1}^{r_f-r_{f,1}+1}H_{f,1}^j\le \Big|s_{f,1}^{r_f-r_{f,1}+1}\Big|F_{f,1}(Z_{f,1}) +\epsilon_{f,1} \\
&= \Big|s_{f,1}^{r_f-r_{f,1}+1}\Big|\Big[\widehat F_{f,1}\big(Z_{f,1}\big|{W}_{f,1}^{*}\big)+\varepsilon_{f,1}(Z_{f,1})\Big]+\epsilon_{f,1}\\
&=\Big|s_{f,1}^{r_f-r_{f,1}+1}\Big|\Big[ {W}_{f,1}^{*}\phi_{f,1}(Z_{f,1})+\varepsilon_{f,1}(Z_{f,1})\Big]+\epsilon_{f,1},
 \end{align*}
where $Z_{f,1}=\big[x_{f,1},x_{l,1,l\in \mathscr {N}_f},x_{l,2,l\in \mathscr {N}_f}\big]^T$, $F_{f,1}=\max\Big\{\big|H_{f,1}^j\big|,j\in\mathpzc M_f\Big\}$, $\epsilon_{f,1}>0$ is a constant and $\varepsilon_{f,1}(Z_{f,1})$ is the approximation error satisfying $\big|\varepsilon_{f,1}(Z_{f,1}) \big| \le \overline {\varepsilon}_{f,1}$ on a compact set $\Omega_{f,1}$, with $Z_{f,1} \in \Omega_{f,1}$ and $\overline \varepsilon_{f,1}>0$ being a constant. The weight $W_{f,1}^{*}$ is the optimal weight vector such that  $W_{f,1}^*=\mathrm{arg}~\underset{{\widehat W_{f,1}^*}}{\mathrm{min}}\bigg\{\underset{\Omega_{Z_{f,1}}}{\mathrm{sup}}\Big| \widehat {F}_{f,1}\big(Z_{f,1}\big|\widehat {W}_{f,1}^{*}\big)- F_{f,1}(Z_{f,1}) \Big|\bigg\}$, with $\widehat {W}_{f,1}^{*}$ being an estimate of $ {W}_{f,1}^{*}$. For subsequent analysis, let us define $\Xi_{f,1}=\big\|W_{f,1}^*\big\|^{\overline {r}_{f,1}}$ and $\Theta_{f,1}=\big\|\phi_{f,1}\big\|$.

Consider the common Lyapunov function candidate
\begin{equation}\label{}
  V_{f,1}=\frac{s_{f,1}^{r_f-r_{f,1}+2}}{r_f-r_{f,1}+2}+\frac{1}{2\beta_{f,1}}\widetilde {\Xi}_{f,1}^2,
\end{equation}
where $\widetilde \Xi_{f,1}=\Xi_{f,1}-\widehat \Xi_{f,1}$.
Using Lemma 1 yields
\begin{align}\label{}
&\Big|s_{f,1}^{r_f-r_{f,1}+1}\Big|F_{f,1}\le \Big|s_{f,1}^{r_f-r_{f,1}+1}\Big|\Big(\big\|W_{f,1}^*\big\|\big\|\phi_{f,1}\big\|+\overline \varepsilon _{f,1}\Big)\notag\\
&\le\frac{1}{\underline {r}_{f,1}}\zeta_{f,1}^{-\underline {r}_{f,1}}+\frac{1}{\overline {r}_{f,1}}\zeta_{f,1}^{\overline {r}_{f,1}}s_{f,1}^{r_f+1}\Big(\big\| W_{f,1}^*\big\| \big\| \phi_{f,1}\big\|\Big)^{\overline r_{f,1}}\notag\\
&~~~~+ \frac{1}{\overline {r}_{f,1}}b_{f,1}^{\overline {r}_{f,1}}s_{f,1}^{r_f+1}+\frac{1}{\underline {r}_{f,1}}b_{f,1}^{-\underline {r}_{f,1}}\overline{\varepsilon}_{f,1}^{~\underline {r}_{f,1}}\notag\\
&\le s_{f,1}^{r_f+1}\bigg(b_{f,1}^{\overline {r}_{f,1}}+\zeta_{f,1}^{\overline {r}_{f,1}}\Xi_{f,1}\Theta_{f,1}^{\overline {r}_{f,1}}\bigg)+\kappa_{f,1},
\end{align}
where $\kappa_{f,1}=\zeta_{f,1}^{-\underline {r}_{f,1}}+b_{f,1}^{-\underline {r}_{f,1}}\overline{\varepsilon}_{f,1}^{~\underline {r}_{f,1}}$ with $\zeta_{f,1}>0$ and $b_{f,1}>0$ being design constants.

In light of (13), (14) and (16), the derivative of $V_{f,1}$ satisfies
\begin{equation}\label{}
\begin{split}
\dot{V}_{f,1}\le& (d_f+\mu_f)s_{f,1}^{r_f-r_{f,1}+1}h_{f,1}x_{f,2}^{r_{f,1}}-\frac{\widetilde \Xi_{f,1}\dot{\widehat \Xi}_{f,1}}{\beta_{f,1}}\\
&+s_{f,1}^{r_f+1}\bigg(b_{f,1}^{\overline {r}_{f,1}}+\zeta_{f,1}^{\overline {r}_{f,1}}\Xi_{f,1}\Theta_{f,1}^{\overline {r}_{f,1}}\bigg)+\hbar_{f,1},
  \end{split}
\end{equation}
where $\hbar_{f,1}=\kappa_{f,1}+\epsilon_{f,1}$. We are now in the position to handle the term $x_{f,2}^{r_{f,1}}$ in (17) through the proposed Lemma 3 as
\begin{align}\label{}
&{s_{f,1}^{r_f-r_{f,1}+1} x_{f,2}^{r_{f,1}}}= s_{f,1}^{r_f-r_{f,1}+1} \big(s_{f,2}+v_{f,1}\big)^{r_{f,1}}\notag\\
 &\le \overline \upsilon_{f,1}\Big| s_{f,1}^{r_f-r_{f,1}+1}s_{f,2}^{r_{f,1}}\Big|  + s_{f,1}^{r_f-r_{f,1}+1}\ell_{f,1}v_{f,1}^{r_{f,1}}.
\end{align}
Then, (17) can be rewritten as
\begin{align}\label{}
\dot{V}_{f,1}\le& (d_f+\mu_f)\overline {h}_{f,1}^j  \overline \upsilon_{f,1}\Big| s_{f,1}^{r_f-r_{f,1}+1}s_{f,2}^{r_{f,1}}\Big| +(d_f+\mu_f)\notag\\
&\times\bigg({h}_{f,1}^j \ell_{f,1}s_{f,1}^{r_f-r_{f,1}+1}v_{f,1}^{r_{f,1}}\bigg) -\frac{1}{\beta_{f,1}}\widetilde \Xi_{f,1}\dot{\widehat \Xi}_{f,1}\notag\\
&+s_{f,1}^{r_f+1}\bigg(b_{f,1}^{\overline {r}_{f,1}}+\zeta_{f,1}^{\overline {r}_{f,1}}\Xi_{f,1}\Theta_{f,1}^{\overline {r}_{f,1}}\bigg)+\hbar_{f,1}.
\end{align}
Substituting the virtual controller $v_{f,1} $ (7) and the adaptation law $\dot{\widehat{\Xi}}_{f,1}$ (12) into (19), and using the fact that
\begin{align}\label{}
   &\overline {h}_{f,1}^j  \overline \upsilon_{f,1}\Big| s_{f,1}^{r_f-r_{f,1}+1}s_{f,2}^{r_{f,1}}\Big|\notag \\
   &\le \overline \tau_{f,1}\bigg(\frac{1}{\overline r_{f,1}}\rho_{f,1}^{\overline r_{f,1}}s_{f,1}^{r_f+1}+\frac{1}{\underline r_{f,1}}\varrho_{f,1}^{-\underline r_{f,1}}s_{f,2}^{r_f+1}\bigg) \notag\\
   &< \overline \tau_{f,1}\bigg(\rho_{f,1}^{\overline r_{f,1}}s_{f,1}^{r_f+1}+\varrho_{f,1}^{-\underline r_{f,1}}s_{f,2}^{r_f+1}\bigg),
\end{align}
we can rewrite (19) as
\begin{align}\label{}
\dot V_{f,1}\le&-c_{f,1}s_{f,1}^{r_f+1}+\big(d_f+\mu_f\big)\overline \tau_{f,1}\rho_{f,1}^{\overline {r}_{f,1}}s_{f,1}^{r_f+1}+\hbar_{f,1}\notag\\
&+\big(d_f+\mu_f\big)\overline {\tau}_{f,1}\varrho_{f,1}^{-\underline {r}_{f,1}}s_{f,2}^{r_f+1}+ \frac{1}{2}\sigma_{f,1}\widetilde \Xi_{f,1}\widehat\Xi_{f,1}\notag\\
\le &-\big(c_{f,1}-\theta_{f,1}\big)s_{f,1}^{r_f+1}+\vartheta_{f,1}s_{f,2}^{r_f+1}+\hbar_{f,1}\notag\\
&+ \frac{1}{2}\sigma_{f,1}\Xi_{f,1}^2- \frac{1}{2}\sigma_{f,1}\widetilde{\Xi}_{f,1}^2,
\end{align}
where $\overline \tau_{f,1}= \overline {h}_{f,1}^j  \overline \upsilon_{f,1}$, $\theta_{f,1}=(d_f+\mu_f)\overline \tau_{f,1}\rho_{f,1}^{\overline r_{f,1}}$ and $\vartheta_{f,1}=(d_f+\mu_f)\overline \tau_{f,1}\varrho_{f,1}^{-\underline r_{f,1}}$ with $\rho_{f,1}>0$ and $\varrho_{f,1}>0$ being design constants.

\textbf{Step $f,2~(f=1,\ldots,N)$ : }Taking the derivative of $s_{f,2}$ yields
\begin{equation}\label{}
  \dot s_{f,2}=h_{f,2}^j(\overline x_{f,2})x_{f,3}^{r_{f,2}}+H_{f,2}^j,
\end{equation}
where $H^j_{f,2}$ is a function defined as
 \begin{align}\label{}
 H_{f,2}^j=&\varphi_{f,2}^j(\overline x_{f,2})-\frac{\partial v_{f,1}}{\partial x_{f,1}}\bigg(\varphi_{f,1}^j(x_{f,1})+h_{f,1}^jx_{f,2}^{{r_{f,1}}}\bigg)\notag\\
 &-\sum_{l\in\mathscr N_f}a_{fl}\frac{\partial v_{f,1}}{\partial x_{l,1}}\bigg(\varphi_{l,1}^j(x_{l,1})+h_{l,1}^jx_{l,2}^{r_{l,1}}\bigg)\notag\\
 &-\frac{\partial v_{f,1}}{\partial y_r}\dot {y}_r-\frac{\partial v_{f,1}}{\partial {\widehat {\Xi}}_{f,1}}\dot {\widehat \Xi}_{f,1},
 \end{align}
Proceeding similarly to Step $f,1$, there exist a continuous function $F_{f,2}\big(Z_{f,2}\big)$ and a linear-in-the-parameter approximator $\widehat F_{f,2}\big(Z_{f,2}\big|{W}_{f,2}^{*}\big)$ such that, for any $j\in\mathpzc M_f$,
 \begin{align*}
&s_{f,2}^{r_f-r_{f,2}+1}H_{f,2}^j\le \Big|s_{f,2}^{r_f-r_{f,2}+1}\Big|F_{f,2}(Z_{f,2}) +\epsilon_{f,2} \\
&= \Big|s_{f,2}^{r_f-r_{f,2}+1}\Big|\Big[\widehat F_{f,2}\big(Z_{f,2}\big|{W}_{f,2}^{*}\big)+\varepsilon_{f,2}(Z_{f,2})\Big]+\epsilon_{f,2}\\
&=\Big|s_{f,2}^{r_f-r_{f,2}+1}\Big|\Big[ {W}_{f,2}^{*}\phi_{f,2}(Z_{f,2})+\varepsilon_{f,2}(Z_{f,2})\Big]+\epsilon_{f,2},
 \end{align*}
where $Z_{f,2}=\Big[\overline {x}_{f,2},\overline {x}_{l,2,l\in\mathscr N_f},\frac{\partial v_{f,1}}{\partial x_{l,1}},\frac{\partial v_{f,1}}{\partial x_{f,1}},\frac{\partial v_{f,1}}{\partial y_r},\frac{\partial v_{f,1}}{\partial {\widehat {\Xi}}_{f,1}}, \widehat \Xi_{f,1},\\ \mu_fy_r \Big]^T$, $F_{f,2}=\max\Big\{\big|H_{f,2}^j\big|,~j\in\mathpzc M_f\Big\}$, $\epsilon_{f,2}>0$ is a constant and $\big|\varepsilon_{f,2}(Z_{f,2}) \big|\le \overline {\varepsilon}_{f,2}$ with $\overline \varepsilon_{f,2}>0$ being a constant. The optimal weight $W_{f,2}^{*}$ and its estimate $\widehat W_{f,2}^{*}$ are defined in a similar way as the previous step. Then, let us define $\Xi_{f,2}=\big\|W_{f,2}^*\big\|^{\overline {r}_{f,2}}$ and $\Theta_{f,2}=\big\|\phi_{f,2}\big\|$.

Consider the common Lyapunov function candidate
\begin{equation}\label{}
V_{f,2}=V_{f,1}+\frac{s_{f,2}^{r_f-r_{f,2}+2}}{r_f-r_{f,2}+2}+\frac{1}{2\beta_{f,2}}\widetilde \Xi_{f,2}^2,
\end{equation}
where $\widetilde \Xi_{f,2}=\Xi_{f,2}-\widehat \Xi_{f,2}$. Along similar lines as (16), we obtain the following inequality
\begin{align}
\Big|s_{f,2}^{r_f-r_{f,2}+1}\Big|F_{f,2}
\le &s_{f,2}^{r_f+1}\Big(b_{f,2}^{\overline {r}_{f,2}}+\zeta_{f,2}^{\overline {r}_{f,2}}\Xi_{f,2}\Theta_{f,2}^{\overline {r}_{f,2}}\Big)\notag\\
&+\kappa_{f,2},
\end{align}
where $\kappa_{f,2}=\zeta_{f,2}^{-\underline {r}_{f,2}}+b_{f,2}^{-\underline {r}_{f,2}}\overline{\varepsilon}_{f,2}^{~\underline {r}_{f,2}}$ with $\zeta_{f,2} > 0$ and $b_{f,2 }> 0$ being design constants. Hence, the derivative of $V_{f,2}$ along (19) and (22) is
\begin{equation*}\label{}
\begin{split}
\dot{V}_{f,2}\le &-\big(c_{f,1}-\theta_{f,1}\big)s_{f,1}^{r_f+1}+h_{f,2}^j(\overline x_{f,2})s_{f,2}^{r_f-r_{f,2}+1}x_{f,3}^{r_{f,2}}\\
&-\frac{1}{\beta_{f,2}}\widetilde \Xi_{f,2}\dot{\widehat \Xi}_{f,2}+s_{f,2}^{r_f+1}\bigg(b_{f,2}^{\overline {r}_{f,2}}+\zeta_{f,2}^{\overline {r}_{f,2}}\Xi_{f,2}\Theta_{f,2}^{\overline {r}_{f,2}}\bigg)\\
&+ \frac{\sigma_{f,1}}{2}\Big(\Xi_{f,1}^2-\widetilde{\Xi}_{f,1}^2\Big)+\vartheta_{f,1}s_{f,2}^{r_f+1}+\hbar_{f,1}+\hbar_{f,2},
\end{split}
\end{equation*}
where $\hbar_{f,2}=\kappa_{f,2}+\epsilon_{f,2}$. Similarly to (18), the use of the proposed Lemma 3 gives
\begin{align}\label{}
  &{s_{f,2}^{r_f-r_{f,2}+1} x_{f,3}^{r_{f,2}}}=s_{f,2}^{r_f-r_{f,2}+1} \big(s_{f,3}+v_{f,2}\big)^{r_{f,2}}\notag\\
& \le \overline \upsilon_{f,2}\Big| s_{f,2}^{r_f-r_{f,2}+1}s_{f,3}^{r_{f,2}}\Big|  + s_{f,2}^{r_f-r_{f,2}+1}\ell_{f,2}v_{f,2}^{r_{f,2}}.
\end{align}

\begin{remark} (departure from state-of-the-art designs)
In order to highlight the distinguishing feature of the proposed design, let us recall the standard designs in  [29], [31]-[35]. There, instead of (26), $x_{f,3}^{r_{f,2}}$ is tackled by subtracting and adding $ v_{f,2}^{r_{f,2}}$, namely,
\begin{align*}
s_{f,2}^{r_f-r_{f,2}+1}x_{f,3}^{r_{f,2}}= s_{f,2}^{r_f-r_{f,2}+1}\bigg[\Big(x_{f,3}^{r_{f,2}}-v_{f,2}^{r_{f,2}}\Big)+v_{f,2}^{r_{f,2}}\bigg].
\end{align*}
Then, the use of Lemmas 1 and 2 yields
\begin{align}\label{}
&s_{f,2}^{r_f-r_{f,2}+1}\Big(x_{f,3}^{r_{f,2}}-v_{f,2}^{r_{f,2}}\Big)\notag\\
  &\le r_{f,2}\bigg|s_{f,2}^{r_f-r_{f,2}+1}\bigg| \big| s_{f,3}\big|\bigg[2^{r_{f,2}-2}\Big(s_{f,3}^{r_{f,2}-1}+v_{f,2}^{r_{f,2}-1}\Big)\notag\\
  &~~~~+\big(s_{f,2}\varsigma_{f,2}\big)^{r_{f,2}-1}\bigg] \le s_{f,2}^{r_f+1}+\overline\varsigma_{f,2}s_{f,3}^{r_f+1},
\end{align}
where $\overline\varsigma_{f,2}=\Big(2^{r_{f,2}-2}r_{f,2}\Big)^{\underline {r}_{f,2}}+\Big(2^{r_{f,2}-2}r_{f,2}\varsigma_{f,2}^{r_{f,2}-1}\Big)^{r_f+1}$. However, for the methods in [29] ,[31]-[35] to work, $ \overline\varsigma _{f,2}$ is incorporated into the virtual control law $v_{f,3}$ to eliminate the extra term $ \overline\varsigma_{f,2}s_{f,3}^{r_f+1}$ $\big($e.g. [31, eq.(5)], [32, eq.(12)], [33, eq.(4)], [34, the equation after (3.11)]$\big)$: this inevitably increases the complexity of the controller structure. It is also worth remarking that the order of the control gain $\overline\varsigma_{f,k}$ in (27) grows dramatically (exponentially) as the order of the subsystems grows, leading to possibly high control gains. This is in contrast with the order of the control gain in (26) which is proportional to the order of the subsystems.
\end{remark}

\begin{remark} (effects of the separation lemma)
The benefits brought by the proposed Lemma 3 can be summarized as: (i) in the first line of (26), the virtual control $v_{f,2}$ can be extracted from $\big(s_{f,3}+v_{f,2}\big)^{r_{f,2}-1}$ directly without involving any inequalities scaling as in $\big($[31, eq.(17)], [32, eq.(29)], [33, eq.(20)], [34, eq.(3.8)]$\big)$, implying that the extra term $ \overline\varsigma_{f,2}$ will not appear in control design and stability analysis; (ii) the term $\upsilon_{f,2}$ in (26) is eventually upper bounded by a constant $\overline \upsilon_{f,2}$, which is independent of $s_{f,3}$ and $v_{f,2}$ and can be easily handled as shown hereafter.
\end{remark}

At this point, similarly to (20), we can bound one of the terms in (26) as
 \begin{align}
 &\overline h_{f,2}^j\overline \upsilon_{f,2}\Big|s_{f,2}^{r_f-r_{f,2}+1}\Big|\Big|s_{f,3}^{r_{f,2}}\Big|= \overline \tau_{f,2}\Big|s_{f,2}^{r_f-r_{f,2}+1}\Big|\Big|s_{f,3}^{r_{f,2}}\Big|\notag\\
 &\le \overline \tau_{f,2}\bigg(\rho_{f,2}^{\overline r_{f,2}}s_{f,2}^{r_f+1}+\varrho_{f,2}^{-\underline r_{f,2}}s_{f,3}^{r_f+1}\bigg),
 \end{align}
where $\overline \tau_{f,2}= \overline {h}_{f,2}^j  \overline \upsilon_{f,2}$, $\rho_{f,2}>0$ and $\varrho_{f,2}>0$ are design constants.

Substituting the virtual controller $v_{f,2}$ (9) and the adaptation law $\dot{\widehat \Xi}_{f,2}$ (12) into the Lyapunov derivative after (25) results in
\begin{equation*}\label{}
\begin{split}
\dot{V}_{f,2}\le& -\big(c_{f,1}-\theta_{f,1}\big)s_{f,1}^{r_f+1}-\big(c_{f,2}-\vartheta_{f,1}-\theta_{f,2}\big)s_{f,2}^{r_f+1}\\
&+\vartheta_{f,2}s_{f,3}^{r_f+1}+\sum_{k=1}^2
\bigg(  \frac{\sigma_{f,k}}{2}\Xi_{f,k}^2-\frac{\sigma_{f,k}}{2}\widetilde \Xi_{f,k}^2+\hbar_{f,k}\bigg),
\end{split}
\end{equation*}
where $\theta_{f,2}=\overline \tau_{f,2}\rho_{f,2}^{\overline r_{f,2}}$ and $\vartheta_{f,2}=\overline \tau_{f,2}\varrho_{f,2}^{-\underline r_{f,2}}$.

\textbf{Step $f,k~(f=1,\ldots,N,~k=3,\ldots,n_f-1)$ : }It follows from (1) and (6) that the derivative of $s_{f,k}$ is
\begin{equation}\label{}
\dot s_{f,k}=h_{f,k}^j(\overline x_{f,k})x_{f,k+1}^{r_{f,k}}+H_{f,k}^j,
\end{equation}
where $H^j_{f,k}$ is a function defined as
 \begin{align}\label{}
 H_{f,k}^j=&\varphi_{f,k}^j(\overline x_{f,k})-\sum_{l\in\mathscr N_f}\frac{\partial v_{f,k}}{\partial x_{l,1}}\Big(\varphi_{l,1}^j(x_{l,1})+h_{l,1}^jx_{l,2}^{r_{l,1}}\Big)\notag\\
 &-\sum_{q=1}^{k-1}\frac{\partial v_{f,k}}{\partial x_{f,q}}\Big(\varphi_{f,q}^j(\overline x_{f,q})+h_{f,q}^jx_{f,q+1}^{{r_{f,q}}}\Big)\notag\\
 &-\sum_{q=1}^{k-1}\frac{\partial v_{f,k}}{\partial \widehat \Xi_{f,q}}\dot {\widehat \Xi}_{f,q}-\frac{\partial v_{f,k}}{\partial y_r}\dot {y}_r.
 \end{align}
Likewise, there exist a continuous function $F_{f,k}\big(Z_{f,k}\big)$ and an approximator $\widehat F_{f,k}\big(Z_{f,k}\big|{W}_{f,k}^{*}\big)$ such that, for any $j\in \mathpzc M_f$,
 \begin{align*}
&s_{f,k}^{r_f-r_{f,k}+1}H_{f,k}^j\le \Big|s_{f,k}^{r_f-r_{f,k}+1}\Big|F_{f,k}(Z_{f,k}) +\epsilon_{f,k} \\
&= \Big|s_{f,k}^{r_f-r_{f,k}+1}\Big|\Big[\widehat F_{f,k}\big(Z_{f,k}\big|{W}_{f,k}^{*}\big)+\varepsilon_{f,k}(Z_{f,k})\Big]+\epsilon_{f,k}\\
&=\Big|s_{f,k}^{r_f-r_{f,k}+1}\Big|\Big[ {W}_{f,k}^{*}\phi_{f,k}(Z_{f,k})+\varepsilon_{f,k}(Z_{f,k})\Big]+\epsilon_{f,k},
 \end{align*}
where $Z_{f,k}=\Big[\overline {x}_{f,k},\overline {x}_{l,2,{l\in\mathscr N_f}}, \frac{\partial v_{f,k-1}}{\partial x_{l,1}},\frac{\partial v_{f,k-1}}{\partial x_{f,1}},\ldots,\frac{\partial v_{f,k-1}}{\partial x_{f,k-1}}$,\\ $\frac{\partial v_{f,k-1}}{\partial \widehat{\Xi}_{f,1}}, \ldots,\frac{\partial v_{f,k-1}}{\partial \widehat{\Xi}_{f,k-1}}, {\widehat {\Xi}}_{f,1},\ldots,{\widehat {\Xi}}_{f,k-1},\frac{\partial v_{f,k-1}}{\partial y_r},\mu_fy_r\Big]^T$, $F_{f,k}$\\$=\max\Big\{\big|H_{f,k}^j\big|,~j\in\mathpzc M_f\Big\}$, $\epsilon_{f,k}>0$ is a constant and $\big|\varepsilon_{f,k}(Z_{f,k}) \big|\le \overline {\varepsilon}_{f,k}$ with $\overline \varepsilon_{f,k}>0$ being a constant. The optimal weight $W_{f,k}^{*}$ and its estimate $\widehat W_{f,k}^{*}$ are defined in a similar way as the previous steps. Let us further define $\Xi_{f,k}=\big\|W_{f,k}^*\big\|^{\overline {r}_{f,k}}$ and $\Theta_{f,k}=\big\|\phi_{f,k}\big\|$.

Consider the common Lyapunov function candidate
\begin{equation}\label{}
V_{f,k}=V_{f,k-1}+\frac{s_{f,k}^{r_f-r_{f,k}+2}}{r_f-r_{f,k}+2}+\frac{1}{2\beta_{f,k}}\widetilde \Xi_{f,k}^2,
\end{equation}
where $\widetilde \Xi_{f,k}=\Xi_{f,k}-\widehat \Xi_{f,k}$. Following similar lines as Step $f,1$ and Step $f,2$, it is possible to obtain the derivative of $V_{f,k}$ as
\begin{align}\label{}
\dot{V}_{f,k}\le&-\sum_{m=1}^k \big(c_{f,m}-\theta_{f,m}-\vartheta_{f,m-1}\big)s_{f,m}^{r_f+1}+\vartheta_{f,k}s_{f,k+1}^{r_f+1}\notag\\
&+\sum_{m=1}^k\bigg(  \frac{\sigma_{f,m}}{2}\Xi_{f,m}^2-\frac{\sigma_{f,m}}{2}\widetilde \Xi_{f,m}^2+\hbar_{f,m}\bigg),
\end{align}
where $\vartheta_{f,0}=0$, $\theta_{f,1}=(d_f+\mu_f)\overline \tau_{f,1}\rho_{f,1}^{\overline r_{f,1}}$, $\vartheta_{f,1}=(d_f+\mu_f)\overline \tau_{f,1}\varrho_{f,1}^{-\underline r_{f,1}}$, $\theta_{f,m}=\overline \tau_{f,m}\rho_{f,m}^{\overline r_{f,m}}$ and  $\vartheta_{f,m}=\overline \tau_{f,m}\varrho_{f,m}^{-\underline r_{f,m}}(m=2,\ldots,k)$, $\overline \tau_{f,m}= \overline {h}_{f,m}^j  \overline \upsilon_{f,m}$ with  $\overline \upsilon_{f,m}$ being the upper bound of $\upsilon_{f,m}(s_{f,m+1},v_{f,m})$, $\hbar_{f,m}=\kappa_{f,m}+\epsilon_{f,m}$, $\kappa_{f,m}=\zeta_{f,m}^{-\underline {r}_{f,m}}+b_{f,m}^{-\underline {r}_{f,m}}\overline{\varepsilon}_{f,m}^{~\underline {r}_{f,m}}$, $\zeta_{f,m}>0$, $b_{f,m}>0$, $c_{f,m}>0$, $\rho_{f,m}>0$ and $\varrho_{f,m}>0$ are design parameters.

\textbf{Step $f,n_f(f=1,\ldots,N)$ :} For the final step, consider the common Lyapunov function candidate
 \begin{align}\label{}
V_{f,n_f}=V_{f,n_f-1}+\frac{s_{f,n_f}^{r_f-r_{f,n_f}+2}}{r_f-r_{f,n_f}+2}+\frac{1}{2\beta_{f,n_f}}\widetilde \Xi_{f,n_f}^2,
 \end{align}
 where $\Xi_{f,n_f}=\big\|W_{f,n_f}^*\big\|^{\overline {r}_{f,n_f}}$ and $\widetilde \Xi_{f,n_f}=\Xi_{f,n_f}-\widehat \Xi_{f,n_f}$.

 Choosing the common actual controller $u_f\triangleq u_f^j$ for the $f$th follower as (11), one immediately gets from (32) that
\begin{align}\label{}
\dot{V}_{f,n_f}
\le&-\sum_{k=1}^{n_f} \big(c_{f,k}-\theta_{f,k}-\vartheta_{f,k-1}\big)\gamma_f^{\frac{{r_{f,k}-1}}{{r_{f}+1}}}s_{f,k}^{r_f-r_{f,k}+2}\notag\\
&+\sum_{k=1}^{n_f}\bigg(  \frac{1}{2}\sigma_{f,k}\Xi_{f,k}^2-\frac{1}{2}\sigma_{f,k}\widetilde \Xi_{f,k}^2+\hbar_{f,k}\bigg)\notag\\
&+\sum_{k=1}^{n_f}\bigg(\gamma_f\big(c_{f,k}-\theta_{f,k}-\vartheta_{f,k-1}\big) \bigg),
\end{align}
where above inequality holds due to $\vartheta_{f,0}=0$, $s_{f,n_f+1}=0$ and the fact that
\begin{equation}\label{}
  \gamma_f^{({r_{f,k}-1})/({r_{f}+1})}s_{f,k}^{r_f-r_{f,k}+2} \le \gamma_f+s_{f,k}^{r_f+1},
\end{equation}
with $\gamma_f>0$ a constant, according to Lemma 1.

\section{Stability analysis}
To analyze the stability of the entire closed-loop system, consider the combined common Lyapunov function
\begin{equation}\label{}
  V=\sum\nolimits_{f=1}^NV_{f,n_f}.
\end{equation}
\begin{theorem}
Under Assumptions 1-3, consider the closed-loop system composed by the high-order switched nonlinear multi-agent system (1), the distributed adaptive consensus controllers (7)-(11) and the parameter adaptation laws (12) with initial conditions $\widehat\Xi_{f,k}(0)\ge 0,~(f=1,\ldots,N,~k=1,\ldots,n_f)$. Then, all signals of the closed-loop system are cooperatively semi-globally asymptotically bounded and the consensus tracking error $\delta$ converges to the following compact set
\begin{equation*}\label{}
 \Omega_3= \left\{\|\delta\|\le   \sqrt {\frac{N^{N-1}\big(N^2+N-1\big)^2\sum_{f=1}^{N}\Big[\frac{\chi}{\overline \alpha}\overline\psi_f\Big]^{\frac{2}{\underline\psi_f}}}{(N-1)^{N-1}}}\right\},
\end{equation*}
where $\overline\psi_f=\mathrm{max}\big\{r_f-r_{f,1}+2,f\in\mathpzc M_f\big\}$, $\underline\psi_f=\mathrm{min}\big\{r_f-r_{f,1}+2,f\in\mathpzc M_f\big\}$, $\overline \alpha$ and $\chi$ are given in the proof.
\end{theorem}

\textbf{PROOF.} See Appendix.

\begin {remark} (comment on practical stability)
Theorem 1 provides the first result available in literature for  practical consensus tracking of high-order nonlinear multi-agent systems with switched dynamics. It is worth pointing out that, as proven in [34], even for a single high-order system, asymptotical tracking is in general impossible (even locally).
\end{remark}

\begin{remark} (design guidelines)
Some guidelines for selecting appropriate design parameters are: (i) choosing small positive constants $\sigma_{f,k}$ and increasing $\beta_{f,k}$ leads to a faster convergence of $\widehat \Xi_{f,k}$; (ii) decreasing $\zeta_{f,k}$, $b_{f,k}$ and $\epsilon_{f,k}$ while increasing $\beta_{f,k}$, and enhancing the connectivity of the communication link contribute to reduce the size of $\Omega_3$.
\end{remark}

\section{Simulation examples}
To validate the effectiveness of the proposed scheme, one leader (labeled by 0) with three follower agents are considered with the directed graph in Fig. 1.
\begin{figure}[h!]
    \centering
\begin{tikzpicture}[scale=1]  

\shade[ball color=red](0,0)circle(0.4cm);
	\node[above right]at(-0.2,-0.2){\textcolor{white}{$0$}};

\shade[ball color=blue](2.4,0)circle(0.4cm);
	\node[above right]at(2.2,-0.22){\textcolor{white}{$1$}};
\shade[ball color=yellow](1.9,1.3)rectangle(1.2cm, 0.6cm);
	\node[above right]at(1.38,0.70){\textcolor{black}{$1$}};
\shade[ball color=yellow](1.9,-1.3)rectangle(1.2cm, -0.6cm);
	\node[above right]at(1.38,-1.14){\textcolor{black}{$2$}};
\shade[ball color=yellow](3.9,0.35)rectangle(3.2cm, -0.35cm);
	\node[above right]at(3.3,-0.22){\textcolor{black}{$3$}};

\shade[ball color=purple](5.4,1.8)circle(0.4cm);
	\node[above right]at(5.2,1.6){\textcolor{white}{$2$}};
\shade[ball color=green](4.9,3.1)rectangle(4.2cm, 2.4cm);
	\node[above right]at(4.3,2.5){\textcolor{black}{$1$}};
\shade[ball color=green](4.9,1.2)rectangle(4.2cm, 0.5cm);
	\node[above right]at(4.3,0.6){\textcolor{black}{$2$}};
\shade[ball color=green](6.9,2.15)rectangle(6.2cm, 1.45cm);
	\node[above right]at(6.3,1.6){\textcolor{black}{$3$}};

\shade[ball color=magenta](5.4,-1.8)circle(0.4cm);
	\node[above right]at(5.2,-2){\textcolor{black}{$3$}};
\shade[ball color=cyan](4.9,-3.1)rectangle(4.2cm, -2.4cm);
	\node[above right]at(4.3,-3){\textcolor{black}{$2$}};
\shade[ball color=cyan](4.9,-1.2)rectangle(4.2cm, -0.5cm);
	\node[above right]at(4.3,-1.1){\textcolor{black}{$1$}};
\shade[ball color=cyan](6.9,-2.15)rectangle(6.2cm, -1.45cm);
	\node[above right]at(6.3,-2.0){\textcolor{black}{$3$}};

\draw[color=red,->,thick](0.4,0)--(2,0);
\draw[color=black,->,thick](2.8,0)--(3.2,0);
\draw[color=black,->,thick](2.2,0.3)--(1.9,0.62);
\draw[color=black,->,thick](2.2,-0.3)--(1.9,-0.62);

\draw[color=red,->,thick](2.65,0.24)--(5.0,1.8);
\draw[color=black,->,thick](5.2,2.1)--(4.9,2.4);
\draw[color=black,->,thick](5.2,1.5)--(4.9,1.2);
\draw[color=black,->,thick](5.8,1.8)--(6.2,1.8);

\draw[color=red,->,thick](5.4,1.4)--(5.4,-1.4);
\draw[color=black,->,thick](5.2,-2.1)--(4.9,-2.4);
\draw[color=black,->,thick](5.2,-1.5)--(4.9,-1.2);
\draw[color=black,->,thick](5.8,-1.8)--(6.2,-1.8);

\draw[color=red,->,thick](5.0,-1.8)--(2.7,-0.27);
\end{tikzpicture}
	\caption{The communication graph between leader 0 and follower agents 1, 2 and 3. Each agent can switch among three dynamics, represented as three squares around each agent.}
	\label{Graph}
\end{figure}
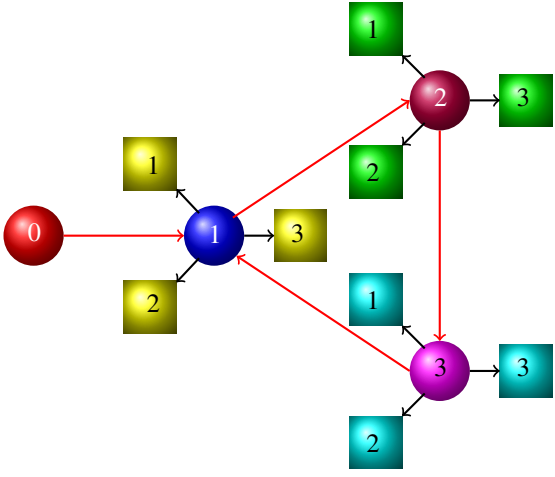
The high-order powers are taken as $r_{1,1}=3$, $r_{1,2}=5$, $r_{2,1}=3$, $r_{2,2}=7$, $r_{3,1}=5$, $r_{3,2}=9$, $n_f=2$, $f=1,2,3$. For each follower, the switching signal is $\sigma_f(\cdot)$: $[0,\infty)\rightarrow\mathpzc M_f=\{1,2,3\}$: note that each follower has its own switching signal, and thus can switch asynchronously with respect to the other followers. The unknown switched nonlinearities $\varphi_{f,k}^{\sigma_f(t)}(\cdot)$ and $h_{f,k}^{\sigma_f(t)}(\cdot)$ are taken to be heterogeneous:\\\\
For follower 1: $\varphi_{1,1}^1=1-\mathrm {cos}(x_1)$, $\varphi_{1,1}^2=0.5+\mathrm{exp}(-x_1^2)$, $\varphi_{1,1}^3=0.2\mathrm{cos}(x_1)+0.5$, $\varphi_{1,2}^1=x_1x_2+0.5$, $\varphi_{1,2}^2=0.2x_1^2+x_2$, $\varphi_{1,2}^3=\mathrm{cos}(x_1^2x_2)+0.2$, $h_{1,1}^1=|\mathrm {tanh}(x_1^2)|+4$, $h_{1,1}^2=\mathrm{cos}(x_1^3)+3$, $h_{1,1}^3=2\mathrm{cos}(x_1)^2$, $h_{1,2}^1=2(|\mathrm{cos}(x_1^3x_2)|+1)$, $h_{1,2}^2=3\mathrm{sin}(x_2)^3+5$, and $h_{1,2}^3=5|\mathrm{sin}(0.1x_1x_2)|+2$. \\\\
For follower 2: $\varphi_{2,1}^1=1.5x_1+x_1^2$, $\varphi_{2,1}^2=x_1^3+0.25$, $\varphi_{2,1}^3=x_1^2+0.2$, $\varphi_{2,2}^1=0.25x_2+0.5$, $\varphi_{2,2}^2=0.3+0.5x_1x_2^2$, $\varphi_{2,2}^3=\mathrm{cos}(x_1x_2)+0.2$, $h_{2,1}^1=2\mathrm{sin}(x_1^2)+6$, $h_{2,1}^2=\mathrm{sin}(x_1^3)+3$, $h_{2,1}^3=\mathrm{cos}(x_1)+3$, $h_{2,2}^1=3\mathrm{cos}(x_2^2)+5$, $h_{2,2}^2=3\mathrm{cos}(x_1+x_2^2)+5$, and $h_{2,2}^3=5+3\mathrm{sin}(x_1^2+x_2x_1^2)$.\\\\
 For follower 3: $\varphi_{3,1}^1=x_1+0.5\mathrm{sin}(x_1)$, $\varphi_{3,1}^2=0.3x_1^2+\mathrm{cos}(x_1)$, $\varphi_{3,1}^3=x_1+0.5x_1^2$, $\varphi_{3,2}^1=0.3x_1^2+x_2$, $\varphi_{3,2}^2=x_2+0.5\mathrm{sin}(x_1)$, $\varphi_{3,2}^3=\mathrm{cos}(x_1x_2)^2+0.5$, $h_{3,1}^1=|\mathrm{cos}(x_1)|+4$, $h_{3,1}^2=|\mathrm{sin}x_2^3|+2$, $h_{3,1}^3=\mathrm{cos}(x_2^2x_1^3)+3$, $h_{3,2}^1=\mathrm{cos}(x_2^2)+3$, $h_{3,2}^2=4\mathrm{cos}(x_1)+6$, and $h_{3,2}^3=\mathrm{cos}(x_2)^3+3$.\\\\
 The leader output is $y_r(t)=2\mathrm{sin}(t)+2\mathrm{sin}(0.5t)$. The initial conditions for the follower agents are taken as: $x_{1,1}(0)=x_{2,1}(0)=x_{3,1}(0)=0.1$, $x_{1,2}(0)=x_{2,2}(0)=x_{3,2}(0)=-0.1$, $\widehat\Xi_{1,1}(0)=\widehat\Xi_{1,2}(0)=5$, $\widehat\Xi_{2,1}(0)=\widehat\Xi_{2,2}(0)=7$ and $\widehat\Xi_{3,1}(0)=\widehat\Xi_{3,2}(0)=10$. The design parameters are chosen to be: $c_{1,1}=c_{2,1}=c_{3,1}=3$, $c_{1,2}=c_{2,2}=c_{3,2}=1.5$, $\beta_{1,2}=\beta_{2,2}=\beta_{3,2}=1$, $\beta_{1,1}=\beta_{2,1}=\beta_{3,1}=15$, $\sigma_{1,1}=\sigma_{2,1}=\sigma_{3,1}=0.5$, $\sigma_{1,2}=\sigma_{2,2}=\sigma_{3,2}=1$, $\zeta_{1,1}=\zeta_{2,1}=\zeta_{3,1}=0.5$, $\zeta_{1,2}=\zeta_{2,2}=\zeta_{3,2}=0.75$, $b_{1,1}=b_{2,1}=b_{3,1}=0.5$ and $b_{1,2}=b_{2,2}=b_{3,2}=1$. The simulation results are shown in Figs. 2-4.  Fig. 2 reveals that the switching signals $\sigma_f(\cdot)$, $f=1,2,3$, for three followers are arbitrary and asynchronous. It can be seen from Fig. 3-(a) that the three followers track the leader signal with bounded consensus tracking errors. Fig. 3-(b) shows the evolutions of control inputs $u_1$, $u_2$ and $u_3$. Fig. 4-(a) and 4-(b) depict the boundedness of $\widehat\Xi_{1,1}$, $\widehat\Xi_{2,1}$ and $\widehat\Xi_{3,1}$, and of $\widehat\Xi_{1,2}$, $\widehat\Xi_{2,2}$ and $\widehat\Xi_{3,2}$, respectively.
\begin{figure}
\centering
\includegraphics[width=3.5 in]{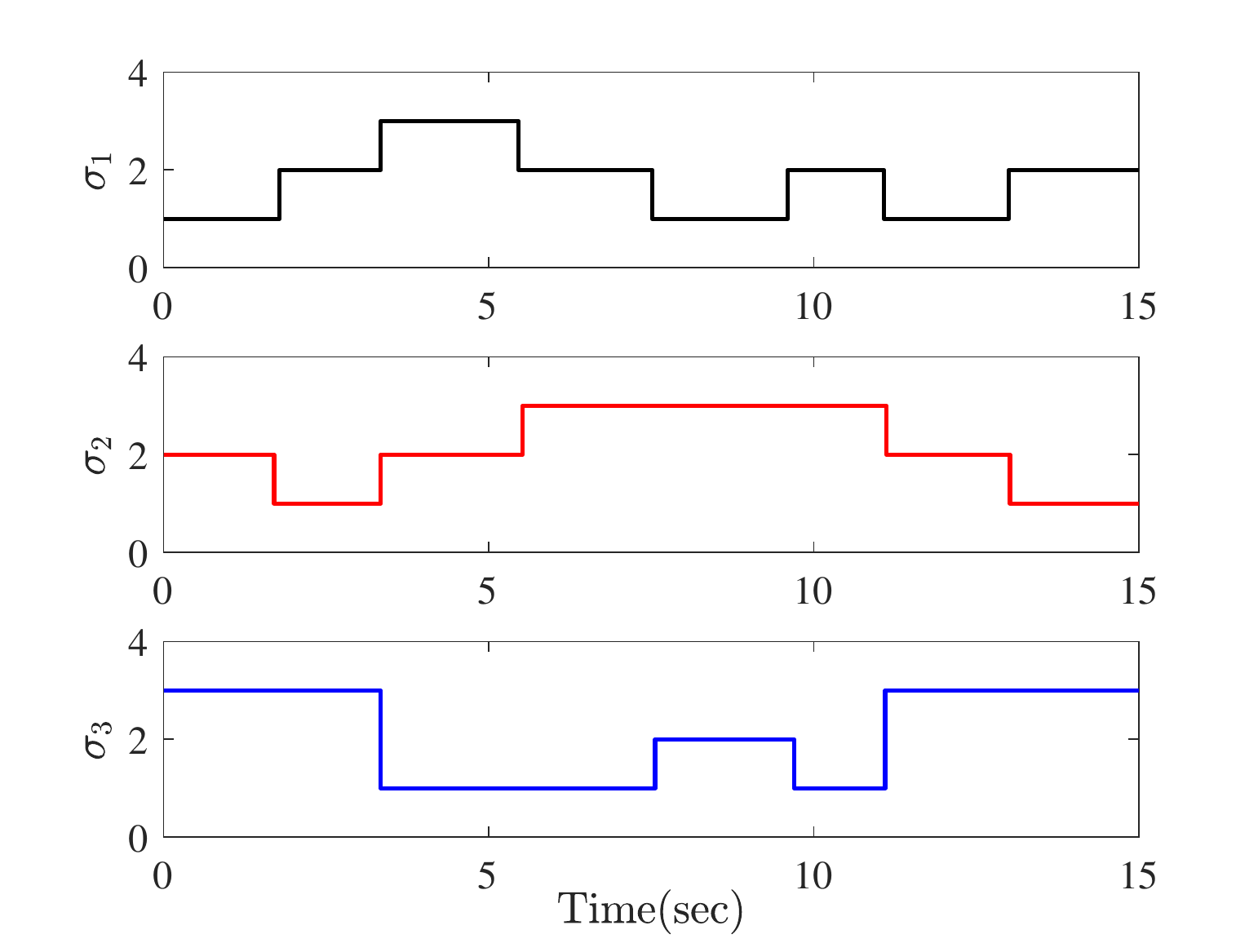}
\caption{Asynchronous switching signals $\sigma_f(t)$.}
\label{}
\end{figure}
\begin{figure}\centering
\includegraphics[width=3.5 in]{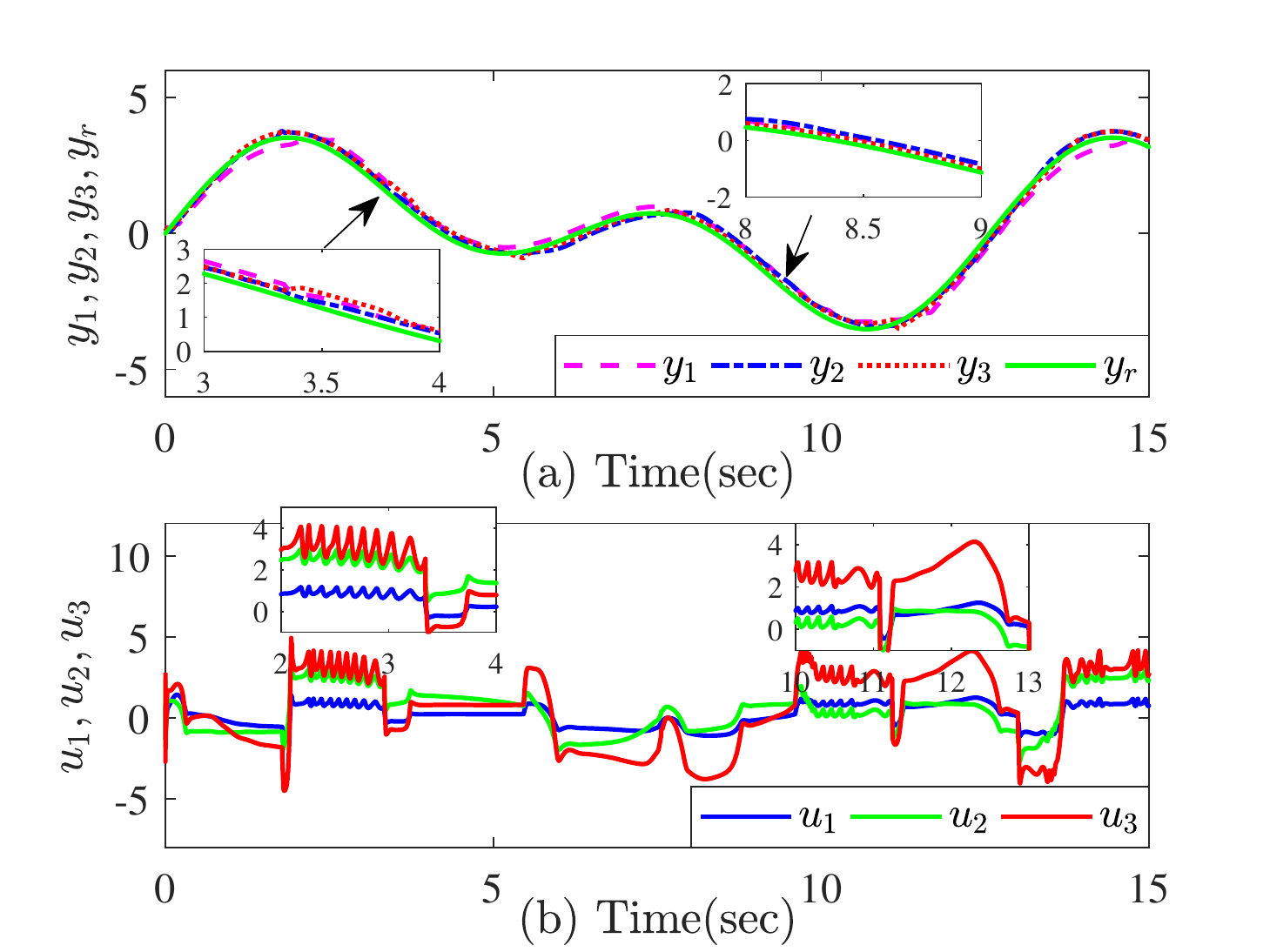}
\caption{(a): Followers outputs $y_1$, $y_2$ and $y_3$, and leader output $y_r$; (b): Control inputs $u_1$, $u_2$ and $u_3$.}
\label{}
\end{figure}\begin{figure}
\centering
\includegraphics[width=3.5 in]{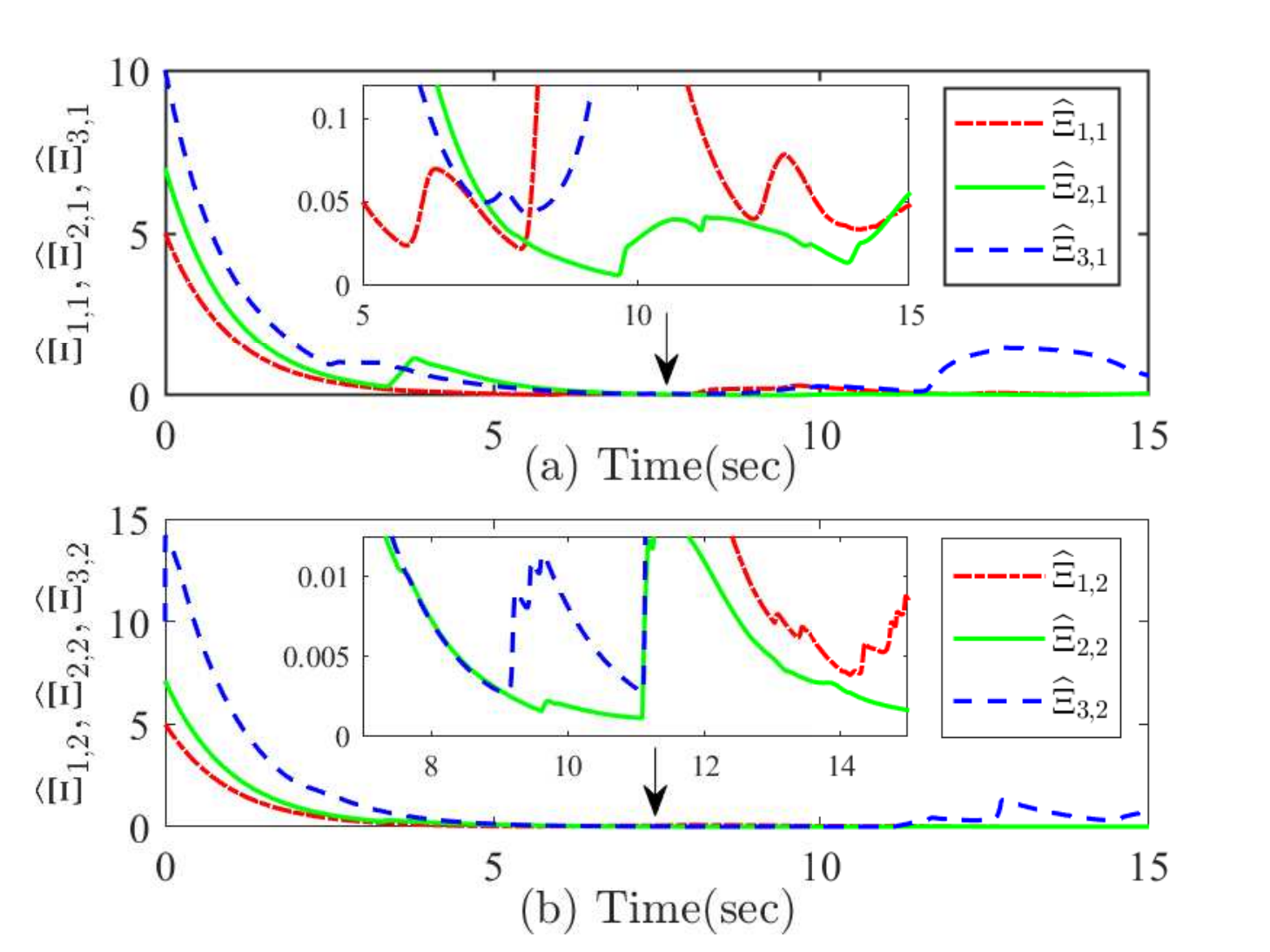}
\caption{The estimates of adaptive parameters of $\widehat\Xi_{1,1}$, $\widehat\Xi_{2,1}$ and $\widehat\Xi_{3,1}$ (a), and of $\widehat\Xi_{1,2}$, $\widehat\Xi_{2,2}$ and $\widehat\Xi_{3,2}$ (b).}
\label{}
\end{figure}
\section{Conclusions}
This paper proposed the first result about distributed consensus tracking for high-order nonlinear multi-agent systems with switched dynamics. This result extends the range of systems for which tracking consensus can be achieved, as existing designs have addressed low-order multi-agent systems. The distinguishing feature of the proposed design is a new separation-based lemma that can simplify the control design in a twofold sense: the complexity of the virtual and actual  control laws is sensibly reduced; the order of the control gains of controllers does not increase exponentially with the order of the subsystems.

\section*{Appendix}
\textbf{Proof of Proposition 1.}
When $x_1\neq0$, without losing generality, we let $x_2=px_1$, $p\in \mathbb R$. Thus, using (ii) yields
\begin{equation}\label{}
  |\digamma(x_1)(\digamma(\overline p)-1)|\le \overline\upsilon(d)|\digamma(p)|\cdot|\digamma(x_1)|+|\digamma(x_1)|d,
\end{equation}
where $\overline p=p+1$. Applying (i) on both sides of (37) gives
\begin{equation}\label{}
  |\digamma(x_1+x_2)-\digamma(x_1)|\le M+|\digamma(x_1)|d,
\end{equation}
where $M=\overline\upsilon(d)|\digamma(x_2)|$. At this point, two situations are considered:\\
\emph{Situation 1}: when $\digamma(x_1)<0$, it follows from (38) that
\begin{equation}\label{}
  \overline d\digamma(x_1)-M\le \digamma(x_1+x_2)\le \underline d\digamma(x_1)+M,
\end{equation}
where $\overline d=d+1$ and $\underline d=1-d$.\\
\emph{Situation 2}: when $\digamma(x_1)\ge0$, one has
\begin{equation}\label{}
  \underline d\digamma(x_1)-M\le \digamma(x_1+x_2)\le \overline d\digamma(x_1)+M.
\end{equation}
When $x_1=0$, (4) becomes
\begin{equation}\label{}
  \digamma(x_2)=\ell(x_1,x_2)\digamma(0)+\upsilon(x_1,x_2)\digamma(x_2),
\end{equation}
which we have to prove. Using (i) we get $\digamma(0)=\digamma(0)\digamma(x_2)$ and (41) becomes $\digamma(x_2)=\big[\ell(0,x_2)\digamma(0)+\upsilon(0,x_2)\big]\digamma(x_2)$ which holds by taking $\ell(0,x_2)\equiv0$ and $\upsilon(0,x_2)\equiv1$. This completes the proof. ~~~~~~~~~~~~~~~~~~~~~~~~~~~~~~~~~~~~~~~~~~~~~~~~~~~~~~~~~~~~~~~~~~~~$\blacksquare$

\textbf{Proof of Lemma 3.} We will verify that condition (ii) in Lemma 3 holds (condition (i) is trivially satisfied). Using the binomial theorem [39, Sect. 3.1, page. 10] leads to
\begin{equation}\label{}
\overline p ^r=1+\frac{p\cdot r!}{(r-1)!}+\cdots+\frac{p^{r-1}\cdot r!}{(r-1)!}+p^r,
\end{equation}
which further results in
\begin{align}\label{}
|\overline p ^r-1|\le \sum_{k=1}^{r}\frac{ r!}{k!(r-k)!}|p|^k\le d+\overline\upsilon(d)| p ^r|,
\end{align}
where the second inequality used Lemma 1, $d=\sum_{k=1}^r\frac{r!}{k!(r-k)!}\frac{r-k}{r}l^{\frac{r}{r-k}}\in (0,1)$ and $\overline\upsilon(d)=\sum_{k=1}^r\frac{r!}{k!(r-k)!}\\
\cdot\frac{k}{r}l^{\frac{-r}{k}}$ with appropriately small constant $l>0$. This completes the proof. ~~~~~~~~~~~~~~~~~~~~~~~~~~~~~~~~~~~~~~~~~~~~~~~~~~~~~~~~~~~~~~~$\blacksquare$

\textbf{Proof of Theorem 1.} It follows from (34) that
\begin{equation*}
  \dot V_{f,n_f}\le -\alpha_fV_{f,n_f}+\varpi_f
\end{equation*}
where $\alpha_f=\mathrm{min}\big\{\big(r_f-r_{f,k}+2\big)\zeta_{f,k},\beta_{f,k}\sigma_{f,k}:1\le k\le n_f\big\}$ with $\zeta_{f,k}= \gamma_f^{({r_{f,k}-1})/({r_{f}+1})}\big( c_{f,k}-\theta_{f,k}-\vartheta_{f,k-1}\big)$ and $\varpi_f=\sum_{k=1}^{n_f}\Big[  \frac{\sigma_{f,k}}{2}\big(\Xi_{f,k}^2-\widetilde \Xi_{f,k}^2\big)+\hbar_{f,k}
+\gamma_f(c_{f,k}-\theta_{f,k}-\vartheta_{f,k-1}) \Big]$. Therefore, the derivative of $V$ can be obtained as
  \begin{equation*}
    \dot V\le -\overline \alpha V+\chi
  \end{equation*}
  where $\overline \alpha= {\mathrm{min}}_{1\le f\le N}\{\alpha_f\}$ and $\chi=\sum_{f=1}^N\varpi_f$. At this point, following similar lines to [31]-[33], it follows that all closed-loop signals remain bounded. From Definition 2, it can be further concluded that the consensus tracking errors of the total closed-loop system are cooperatively semi-globally asymptotically bounded.

  A bound on the tracking error can be obtained as follows: integrating $\dot V(t)$ on $[0,t]$ gives
  \begin{equation*}
    \int_0^t\mathrm {d}\big[\exp(\overline \alpha t)V(t)\big]\le \int_0^t \chi \exp(\overline \alpha t)~\mathrm {dt}
  \end{equation*}
  which suggests that
  \begin{equation*}
    V(t)\le \Big(V(0)-\frac{\chi}{\overline \alpha}\Big)\exp(-\overline \alpha t)+\frac{\chi}{\overline \alpha}
  \end{equation*}
and ${\mathrm{lim}}_{t\rightarrow +\infty}V(t)\le \frac{\chi}{\overline \alpha}$. Thus, one can conclude that ${\mathrm{lim}}_{t\rightarrow +\infty}\frac{s_{f,1}^{r_f-r_{f,1}+2}}{r_f-r_{f,1}+2}\le \frac{\chi}{\overline \alpha}$, leading to
  \begin{align*}
{\mathrm{lim}}_{t\rightarrow +\infty}\|s_1\|\le \sqrt {\sum_{f=1}^N\Bigg[\bigg(\frac{\chi}{\overline \alpha}\overline\psi_f \bigg)^2\Bigg]^{\frac{1}{\underline\psi_f}}}=\Gamma.
  \end{align*}
Then, from the inequalities below (5), one gets that ${\mathrm{lim}}_{t\rightarrow +\infty}\|\delta\|\le \frac{\Gamma}{\underline{\lambda}_{\mathrm{min}}(\overline {\mathscr L}+\mathscr B)}$. Note that $\underline{\lambda}_{\mathrm{min}}\big(\overline{\mathscr L}+\mathscr B\big)$ might not be directly known in distributed control because it is a global topology variable. To handle this issue, one possible solution is to replace $\underline{\lambda}_{\mathrm{min}}\big(\overline{\mathscr L}+\mathscr B\big)$ by a more conservative bound $\frac{\overline N}{N^2+N-1}$ with $\overline N=\big(\frac{N-1}{N}\big)^{\frac{N-1}{2}}$ [40], which only depends on the number of agents. This concludes the proof.$\blacksquare$

\section{Conclusions}
This paper proposed the first result about distributed consensus tracking for high-order nonlinear multi-agent systems with switched dynamics. This result extends the range of systems for which tracking consensus can be achieved, as existing designs have addressed low-order multi-agent systems. The distinguishing feature of the proposed design is a new separation-based lemma that can simplify the control design in a twofold sense: the complexity of the virtual and actual  control laws is sensibly reduced; the order of the control gains of controllers does not increase exponentially with the order of the subsystems.



\end{document}